\documentclass[12pt,preprint]{aastex}

\shorttitle{ R~Aquarii outer jets}
\shortauthors{Kellogg et al}

\begin{document}


\title{Outer jet X-ray and radio emission in R~Aquarii: 1999.8 to 2004.0}


\author{E. Kellogg, C. Anderson, K. Korreck, J. DePasquale, J. Nichols
and J.~L. Sokoloski\footnote{NSF Astronomy \& Astrophysics Fellow.}} \affil{Harvard/Smithsonian Center for Astrophysics,
60 Garden St. Cambridge, MA  02138}

%%\author{J. Pedelty\altaffilmark{1}}

\author{M. Krauss} \affil{ Kavli Institute for Astrophysics and Space
Research, MIT, Cambridge MA 02139}

\author{J. Pedelty}

\affil{Planetary Systems Laboratory, Code 693, NASA Goddard Space
 Flight Center, Greenbelt, MD 20771} \email{ekellogg@cfa.harvard.edu}


%%\altaffiltext{1}{Absentee Professor, University of Maryland, Baltiwash.}


\begin{abstract}
{\it Chandra} and VLA observations of the symbiotic star R~Aqr in 2004
reveal significant changes over the three to four year interval
between these observations and previous observations taken in with the
VLA in 1999 and with {\it Chandra} in 2000.  This paper reports on the
evolution of the outer thermal X-ray lobe-jets and radio jets.  The
emission from the outer X-ray lobe-jets lies farther away from the
central binary than the outer radio jets, and comes from material
interpreted as being shock heated to $\simeq$ 10$^{6}$K, a likely
result of collision between high speed material ejected from the
central binary and regions of enhanced gas density.  Between 2000 and
2004, the Northeast (NE) outer X-ray lobe-jet moved out away from the
central binary, with an apparent projected motion of $\simeq$ 580 km
s$^{-1}$. The Southwest (SW) outer X-ray lobe-jet almost disappeared
between 2000 and 2004, presumably due to adiabatic expansion and
cooling. The NE radio bright spot also moved away from the central
binary between 2000 and 2004, but with a smaller apparent velocity
than of the NE X-ray bright spot.  The SW outer lobe-jet was not
detected in the radio in either 1999 or 2004.  The density and mass of the
X-ray emitting material is estimated.Cooling times, shock
speeds, pressure and confinement are discussed.
\end{abstract}




\section{Introduction}

R Aquarii is a well studied symbiotic system made up of a Mira-type
giant star and a compact object, most likely a white dwarf.  The
system has been observed in the radio \citep{kaf89}, optical
\citep{sol01,ho91,sol02,pa01,ho97}, UV \citep{ho91}, and X-rays
\citep{kel01}, hereafter KPL. The R Aqr system was first reported as
an unresolved {\em Einstein} X-ray source by \citet{ju01}. It was later
observed with {\em EXOSAT} \citep{vi01} and {\em ROSAT}
\citep{hu01}. Since this system is relatively close, about 200 pc away
\citep{ho97a}, there has been intense study of its evolution.  R
Aquarii is contained in an inner nebula extending
$\sim$1\arcmin\,north-south and another that is
$\sim$2\arcmin\,east-west.  One of the most interesting features of
the system is a bi-polar collimated outflow first observed around 1977
\citep{wal80, her80}.

Collimated non-relativistic outflows have been found in accreting
compact stars -- symbiotic stars and supersoft X-ray sources. They
have also been observed in Herbig-Haro objects. X-ray jets may also be
produced in the fast collimated winds from the central stars or
binaries in some bipolar planetary nebulae \citep{sok01}, and
recently, a possible X-ray jet has been reported in the T Tauri star
DG Tau \citep{gud}, and optical spectroscopic evidence for jets in Hen
3-1341 was reported by \citet{to00}. Also see Table 2 of \citet{br04}.

The symbiotic star R~Aqr is the most dramatic example of a
white-dwarf collimated outflow, and it was the first white dwarf found
to have an X-ray jet (KPL). There are currently only two white dwarfs
known to have X-ray jets, R~Aqr and CH Cyg \citep{gal01}, and both are
in symbiotic stars. In both cases, X-rays are present due to shock
heating of the outflowing material as it collides with surrounding
nebular material. The jet in CH Cyg is not completely resolved in the
{\em Chandra} observation, but is visible in an HST observation shown
in the same publication. It is more similar to the central R~Aqr
source and inner jets \citep{ni01} than to the outer lobe-jets discussed
here. It is also possible that the CH Cyg jet spectrum contains an
admixture of emission from the central CH Cyg source because the jet
is not completely spatially resolved. That could explain the counts at
~5 keV from the jet - they could be from the central source, as seen
in R Aqr.

Supersoft X-ray sources (SSS) show evidence for jet outflow but jets
like those we observe in R Aqr have not been imaged directly. This is
not surprising because most SSS are much more distant, being at
kiloparsec distances in our galaxy, in the Magellanic Clouds, or even
in external galaxies, making their angular sizes likely too small to
be resolved. The primary evidence for jets in SSS is the observation
of pairs of red and blue shifted ``satellite'' optical emission lines
in their spectra \citep{co01,ka01}. SSS are typically close binaries
and are far more luminous than R Aqr in X-rays, with L$_x$ in the
near-Eddington range 10$^{37-38}$ erg s$^{-1}$. The radial velocities
of these jets range up to thousands of km s$^{-1}$, considerably
higher than the outer lobe-jet R Aqr velocity. These SSS jets are
thought to be radiation driven. However, they are likely to be
situated very close to the WD accretion disk, whereas the R Aqr lobe-jets
we study here are very far away from the central source, with an expectation of lower velocity due to various sources of drag or dissipation.

Production of X-rays from non-relativistic jets occurs in some
protostellar jets in Herbig-Haro objects (e.g.
\citet{pra01} in which there is marginal evidence for extended
emission in HH2 below 2 keV, but not enough detail to know if there is
a jet. An expanding x-ray feature was seen in HH154, moving outward at
$\sim500$ km s$^{-1}$ \citep{fav02,fa06}. These jets in YSOs, the
Herbig-Haro (HH) objects, are similar to those seen in R Aqr. HH
objects have been known for 50 years to be luminous condensations of
gas in star-forming regions, but not fully explained. A jet-induced
shock model is favored, with material streaming out of a young stellar
object, colliding with the ambient medium. HH2 was the first HH object
discovered as an X-ray source by \citet{pra01}. Eleven photons were
detected, establishing that the source was extended by about 1-2 arc
sec.  The position of the emission is known to 1/2 arc second accuracy
and is at the leading edge of an H$_\alpha$ feature moving at several
hundred km s$^{-1}$ away from the star of origin, an obscured YSO
named VLA 1. The shape of the spectrum cannot be determined from so
few counts, but the counts are all at energies $<$2 keV, similar to
the R Aqr outer lobe-jets. HH154 was observed by \citet{fa06} in 2001
and 2005.

The Herbig Ae star HD 163296 is also an X-ray source with a suggestion
of emission in the direction of a Ly$_\alpha$ emitting jet and
Herbig-Haro outflow \citep{sw01}. The authors fit variable abundances
with elements of similar first ionization potentials fixed at their
relative solar abundances to allow for chemical fractionation commonly
observed in late-type stars of this type.  Most of the X-ray emission
comes from the central star, and is fitted to a mekal thermal plasma
emission model with kT = 0.49 $\pm$ 0.03 keV, a somewhat higher
temperature than for the outer lobe-jets in R Aqr, and qualitatively
different from the central R Aqr source
\citep{ni01}. The suggestion of X-ray emission from a jet was only
five counts, so not much could be learned, except that the counts are
at about 0.8 keV, considerably higher energy than for the R Aqr outer
lobe-jets.

The remainder of this article gives a description of the 2004.0 {\it
Chandra} and VLA observations of the outer lobe-jets in R Aqr, comparing
with the earlier 2000.7 ({\it Chandra} and 1999.8 (VLA) epochs,
presenting images and contour maps, and giving results on X-ray
spectra. The X-ray data, including image and spectral analysis, are
presented in \S\ref{sec:chandraobs}.  The radio data are presented in
\S\ref{sec:vlaobs}. In \S\ref{sec:interp}, we discuss the expected
vs. observed cooling times, the mass, mass loss rate and kinetic
energy in the lobe-jets, and the inferred shock speeds.  We conclude
with a summary of our results in \S\ref{sec:concl}.

\section{Observations and Results}

%% In this section, we use  the \subsection command to set off
%% a subsection.  \footnote is used to insert a footnote to the text.


%% This section also includes several of the displayed math environments
%% mentioned in the Author Guide.

\subsection{{\it Chandra} Observations\label{sec:chandraobs}}

The new X-ray observations reported here were taken with {\it Chandra}
on  2003 December 31 15:05:40 UT, MJD 53004.62894 start time
(hereafter 2004.0) with the ACIS-S3 back illuminated detector
\citep{cxc01}. The dataset is sequence 300140 and obsid 4546
\footnote{available at
\dataset{http://cfa.harvard.edu/chaser/mainEntry.do}} with an exposure
time of 36523~s.

In order to compare the obsid 651 observation taken in 2000.7 with the
new data, we reprocessed the 651 data to apply the same calibration
and processing techniques to both observations (see cxc.harvard.edu
for a description of the current calibration system).  Our
reprocessing of the data used CALDB version 2.28, and the most current
correction algorithms as of Aug. 11, 2004.  We applied a
time-dependent gain correction and the CTI correction to the original
level~1 event file, and created new rmf's for each observation.  We
removed pixel randomization, and then applied the sub-pixel
repositioning algorithm\citep{Li01}.  These new level~1 event files
were then filtered by grade and status, rejecting inappropriate
events, to create level~2 event files, used in the analysis.

\subsubsection{X-ray images\label{sec:xrayim}}

\begin{figure}[h]
%%\hfill
%%\includegraphics[angle=0,scale=.6]{f1.eps}
\plotone{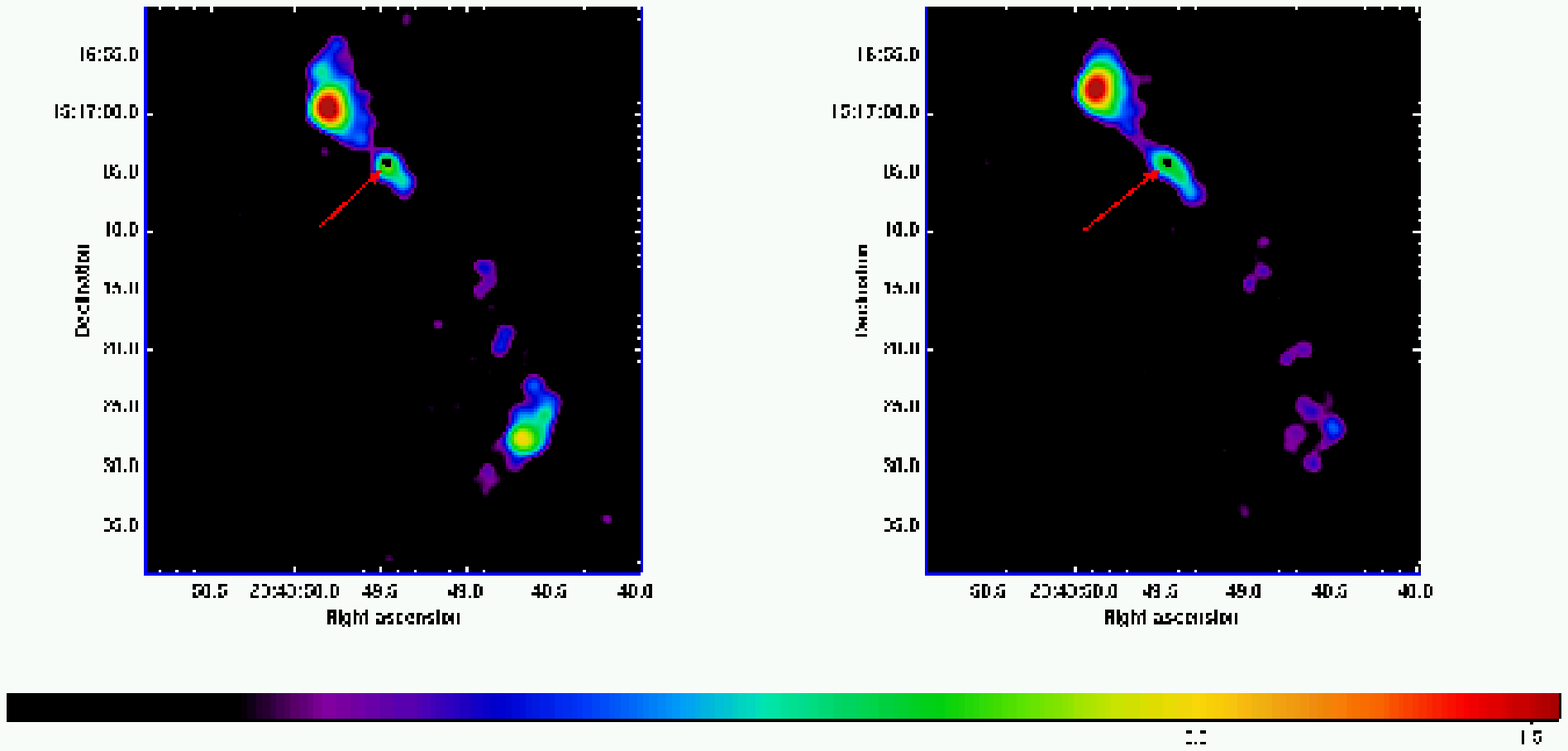}

\caption{{\it Chandra} X-ray images of R~Aqr, 2000.7 above, 2004.0 below. This smoothed false color image represents X-ray counts in the
energy range 0.2-3.5 keV. North is up, west is right. The arrow points to the
central R Aqr binary at 23$^{\rm h} 43^{\rm m} 49\fs45 \,-15 \arcdeg \, 17\arcmin \,
4\farcs2$ J2000, shown as a black dot; color bars are log intensity, in units of counts per 1/8$\arcsec$ pixel. Between 2000.7 and
2004.0 the peak of the NE outer X-ray lobe-jet moved 2$\arcsec$ to the
northwest, away from the central binary, and the SW outer lobe-jet faded
significantly. \label{2epXX}}

\end{figure}

\begin{figure}[h]
%%\hfill
%%\includegraphics[angle=0,scale=.8]{f2.eps}
\plotone{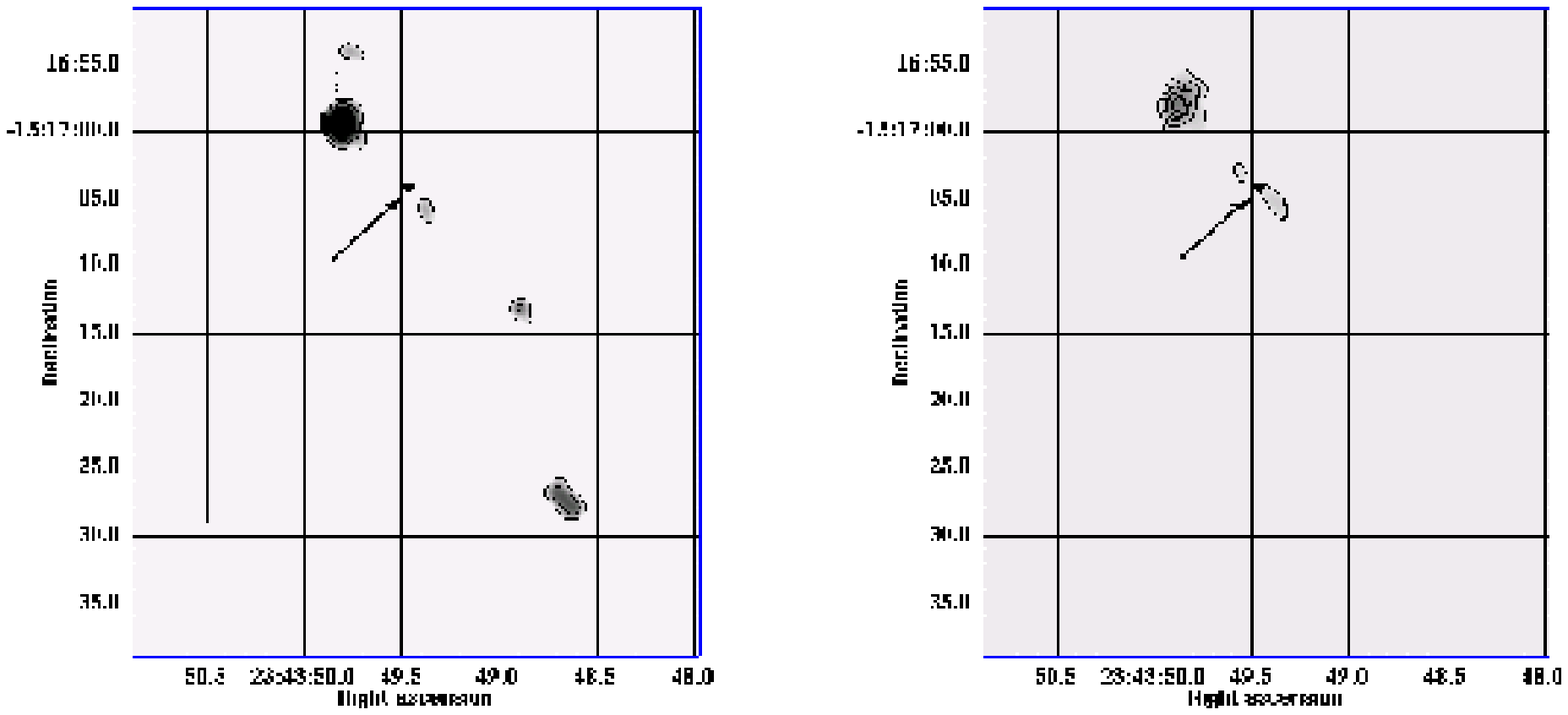}
\caption{{\it Chandra} X-ray images of R~Aqr in the 0.30 keV energy
range. 2000.7 is above, 2004.0 below. North is up, west is right. The arrow points to the location of the central R Aqr
binary; the grayscale bar is log intensity, in units of counts per 1/8$\arcsec$ pixel. \label{2epXXred}}
\end{figure}

\begin{figure}[h]
%%\hfill 
%%\includegraphics[angle=0,scale=.8]{f3.eps}
\plotone{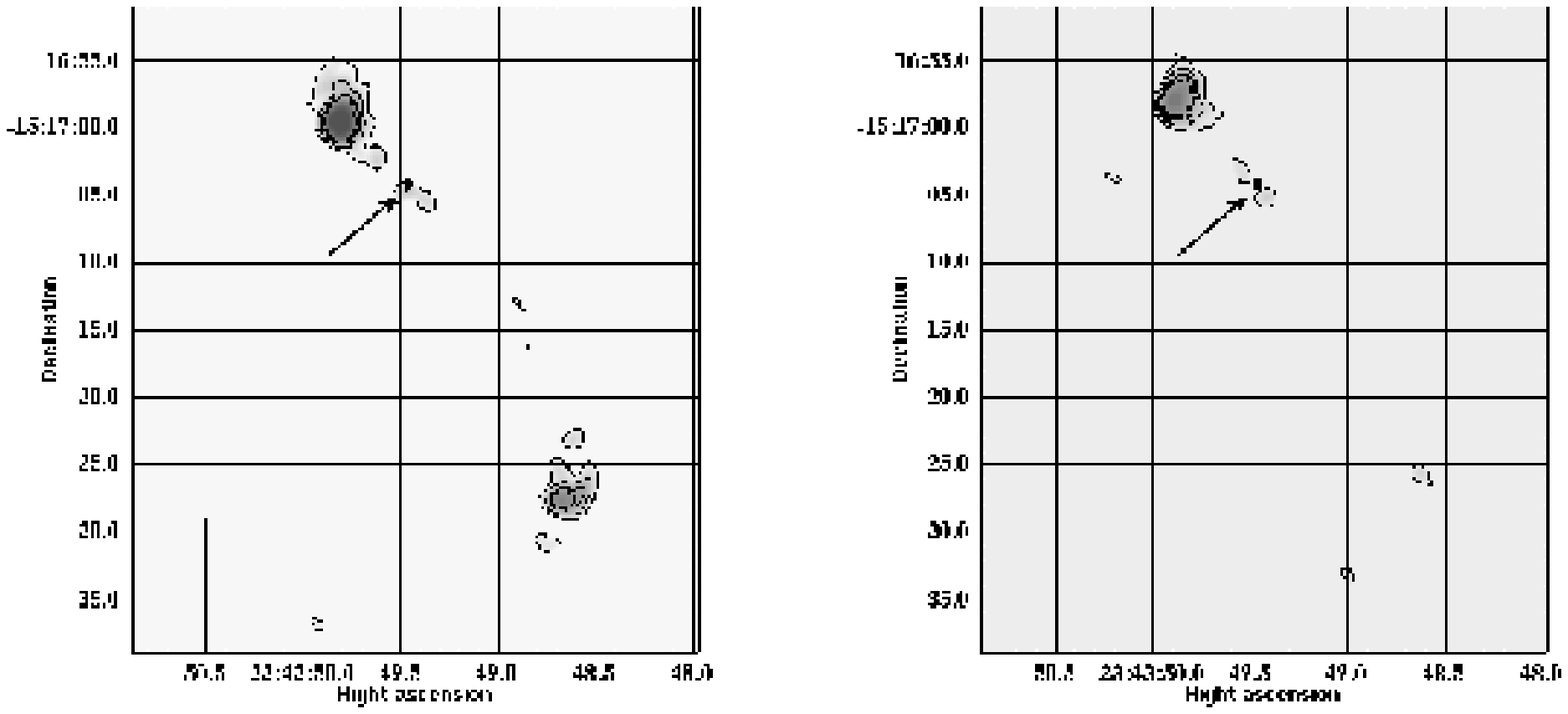}
\caption{{\it Chandra} X-ray images of R~Aqr from the 0.43 keV energy range.
2000.7 is above, 2004.0 below.  North is up, west is right.  The arrow points to the location of the central R Aqr
binary; the grayscale bar is log intensity, in units of counts per 1/8$\arcsec$ pixel.\label{2epXXgreen}}
\end{figure}

\begin{figure}[h]
%%\hfill 
%%\includegraphics[angle=0,scale=.8]{f4.eps}
\plotone{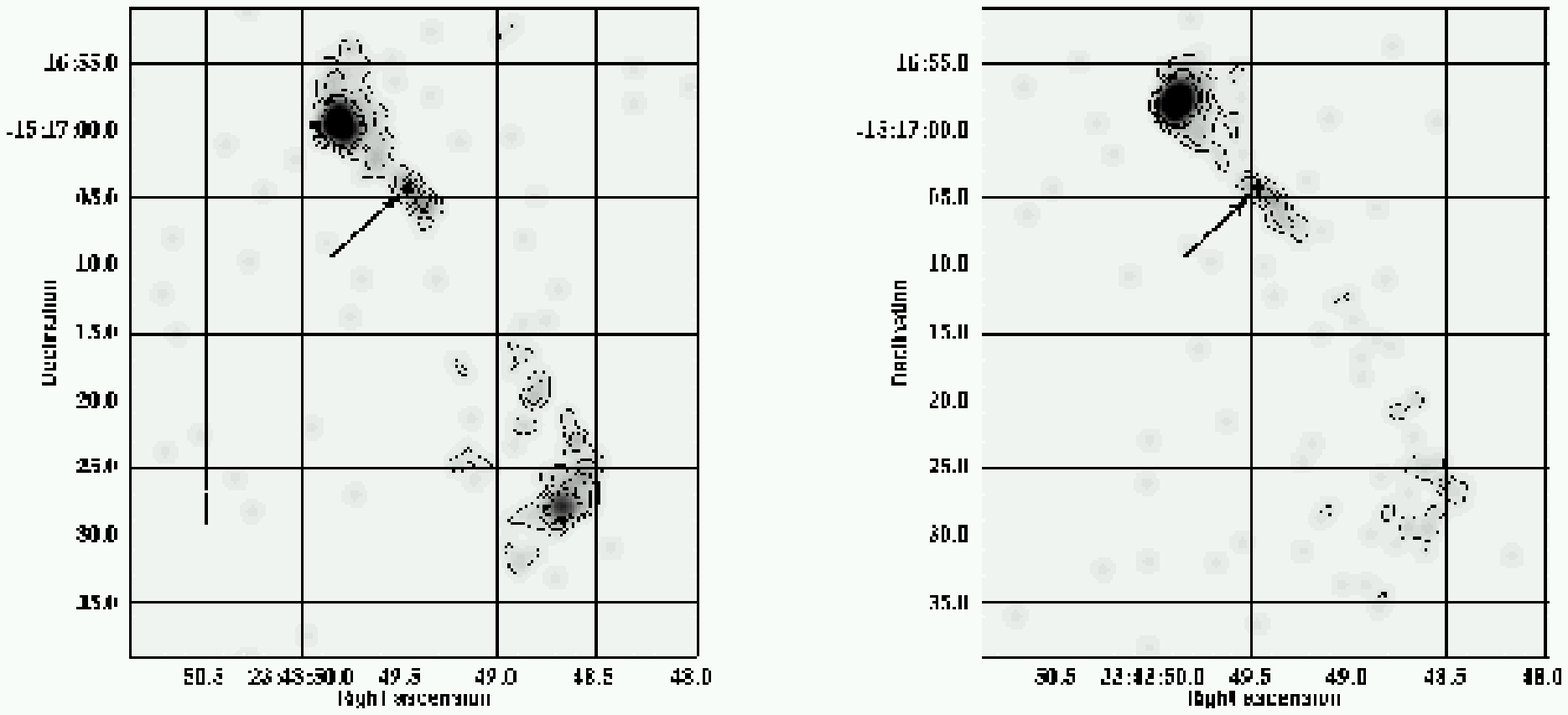}
\caption{{\it Chandra} X-ray images of R~Aqr from the O~VII energy range, $\sim$0.57 keV.
2000.7 is above, 2004.0 below.  North is up, west is right.  The arrow points to the location of the central R Aqr
binary; the grayscale bar is log intensity, in units of counts per 1/8$\arcsec$ pixel. \label{2epXXblue}}
\end{figure}

Figures  \ref{2epXX}, \ref{2epXXred}, \ref{2epXXgreen} and
\ref{2epXXblue}  show a general X-ray view of R~Aqr in the recent
2004.0 epoch, compared with the previous 2000.7 epoch. The total
number of counts in 2004.0 with E$\le$8~keV from a 0.083 square arc
min area encompassing the region in these figures is about 1000, or
0.027 counts~s$^{-1}$. The total counts from a region of the same size
nearby, an indication of background, is 47, or 0.0013
counts~s$^{-1}$.

The images in figures \ref{2epXX}, \ref{2epXXred}, \ref{2epXXgreen}
and \ref{2epXXblue} were created by first resampling the level~2 event
files to 0.25 ACIS pixels, so the resulting pixel size is 1/8$\arcsec$.  A smoothing algorithm available in CIAO,
aconvolve, was used with a 1-$\sigma$ gaussian profile one ACIS pixel
in extent to create the final images.  This removes graininess without
introducing significant blurring.  The false color representation in
Figure \ref{2epXX} has colors assigned on the basis of source
brightness, with a color bar attached.Although the two observations were of different exposure times, they were adjusted for decreasing efficiency of the ACIS-S3 detector so equal numbers of counts in this 0.2-3.5 keV energy range would be expected if the source strength didn't vary.

The X-ray morphology of R Aqr consists of several distinct structures.
X-ray emission is detected from the central interacting binary, and
from a small inner jet to the SW within $\sim 5\arcsec$ of the central
binary (clearly present for the first time in 2004.0; see \citet{ni01}
for more discussion of this new X-ray jet), an outer jet to the NE and
an outer jet to the SW.  While the inner jet and the outer NE lobe-jet are
largely aligned along a line at PA$\simeq 46\degr$, the peak of the
outer SW lobe-jet is located at PA$\simeq 211\degr$, a significant
difference from the position angle of $226\degr$ expected from a
simple opposing jet.  In general, the SW outer lobe-jet is fainter and more
extended than the NE lobe-jet.  This kind of change in position angle has
been interpreted as evidence for precession of the accretion disk
close to the binary \citep{kaf86}. Alternately, we suggest the
jet may be encountering a wall of material constraining its motion,
such as described by \citet{sok01} and \citet{sol02}.

In 2000.7, the NE outer lobe-jet emission was dominated by a bright
spot approximately 8$\arcsec$ from the central binary.  The NE
lobe-jet also produced diffuse emission extending all the way from the
central binary out to a small, northward-curving spur-like structure
roughly 12$\arcsec$ away (see Fig.~\ref{2epXX}). In 2004.0, the NE
outer lobe-jet bright spot had moved outward by $2.0\arcsec
\pm0.1\arcsec$.  This change in position corresponds to an apparent
velocity of $580d \pm 30d$ ~km s$^{-1}$ in the plane of the sky, where
{\it d} is the distance to R Aqr divided by 200 pc.  The extent of the
more diffuse emission, however, remained approximately the same.
\citet{bod98} did a specrocoptic optical observation of R Aqr and
observed outward motion of the same region as our NE lobe-jet in
H$_\alpha$ with a tangential velocity of 550$\pm$300 km s$^{-1}$,
agreeing with our measurement within the errors.

The total number of counts from the NE outer lobe-jet was about the
same in 2000.7 and 2004.0.  Taking into account the the decrease in
effective area resulting from contamination buildup on the ACIS-S3
optical blocking filter and increased exposure time, there was also no
statistically significant change in total flux from the NE outer
lobe-jet between the two epochs (see Table \ref{netbl}). The constancy
of the NE lobe-jet flux is remarkable given that the morphology of the
source changed significantly between the two observations. If the
nominal change were truly occurring, it would be consistent with a
factor of five reduction in flux in the ten years between the
1990-1991 observation \citep{hu01} and our 2000.7 epoch
observation. However, the \citet{hu01} observation with {\em ROSAT}
PSPC could not resolve the NE and SW lobe-jets and the central source
from each other, and did not have spectral resolution nor range
comparable with our {\em Chandra} observations, so more detailed
conclusions cannot be drawn.

In 2000.7, the SW outer lobe-jet consisted of a series of
disconnected, faint emission regions (with a slightly southward
curvature) and a bright spot 26$\arcsec$ from the central binary.  The
SW outer lobe-jet bright spot was $\sim$3 times fainter than the NE
outer lobe-jet bright spot. In 2004.0, the appearance of the SW outer
lobe-jet had changed dramatically.  The SW outer lobe-jet X-ray
emission faded by 75\% overall (see Table \ref{swremtbl}).  The
disappearance of the X-ray bright spot at the center of the SW
lobe-jet (see Figure \ref{2epXX}) accounted for much of this change,
as it was undetectable in 2004.0 The other SW outer lobe-jet emission
regions did not fade significantly, and some may have moved outward
compared to their location in 2000.7, but two images in $\sim$3 years
are inadequate to properly track these possible motions.

Figures \ref{2epXXred}, \ref{2epXXgreen} and \ref{2epXXblue} are based
on our analysis of the energy spectrum of the outer lobe-jets, which
are best modeled by a thermal emission process.the figures show X-rays
in three energy ranges corresponding to expected dominant peaks in
line emission from a thermal plasma, the Si~XI complex just below the
C absorption edge at 0.284, N VI, at$\sim$0.43 and O VII, at 0.57 keV
\citep{zom}.  The contour levels in figures \ref{2epXXred},
\ref{2epXXgreen} and \ref{2epXXblue} are 0.01, 0.025, 0.05, 0.065,
0.1, 0.6 counts pixel$^{-1}$.

Several interesting features are apparent. In Figure \ref{2epXXred}
the 0.3 keV emission in the SW lobe-jet seems to come only from the center
of that region. It may be cooler than the outer part.
Of note in Figure \ref{2epXXgreen} is the region to the north of the
brightest part of the lobe-jet NE of the visible star in the 2000.7
epoch. This suggests it is emitting strongly in this part of
the spectrum. In the 2004.0 epoch, the brightest part of the NE lobe-jet
has moved into that region.  The 0.3
keV part of the SW outer lobe-jet has faded considerably in 2004.0 but not
entirely.  Figure \ref{2epXXblue}, the 0.57 keV range, shows the bulk
of the emission from the outer lobe-jets. It also shows the north spur in
the earlier observation where the later emission has moved. There
is a noticeably larger region of emission in this energy range in the
outer SW lobe-jet, and the center of the SW lobe-jet disappears in the later
observation.

\subsubsection{X-ray spectra}

The spectral data were extracted from data we processed to give the
most up to date results, yielding level~2 event files We extracted
pulse height files using the CIAO {\em specextract} tool, with 10 counts per
bin to allow using the $\chi^2$ technique. The spectra were fit with
XSPEC 12.3.0e \citep{Arn96}; also see
http://heasarc.gsfc.nasa.gov/docs/xanadu/xspec. The \citet{and}
abundances were used.

Given the noticeable peaks in the spectrum suggesting line emission,
we had to decide whether to use an equilibrium thermal model in xspec
or a non-equilibrium collisional model (e.g. the nei model in
xspec). With the relatively small number of counts from the outer lobe-jets
in these spectra, we found we could not constrain the parameters of
complex models such as nei effectively, whereas the equilibrium model
gives an acceptable fit to the data.  Given the limited number of
counts, we had to be careful to fit the data with as few parameters as
possible to obtain significant results. With the apparent thermal
shape of the spectrum, we chose the most up to date thermal model,
APEC \citep{sm01}.  We also tried substituting other spectral forms
instead of APEC. Black body and power law models did not give an
acceptable fit. In all the fits we included interstellar absorption
fixed at the 21-cm value of N$_{H}$ = 1.85 $\times$10$^{20}$
\citep{Stark92}. The entries in Table \ref{netbl} entitled 
''Source 0.25-2.0 keV flux'' were computed with the interstellar
absorption set to zero.

In KPL we had applied the more complex non-equilibrium model to
the data; in this work we found the simpler model gave just as
acceptable a fit to the data without the model complexity. We believe
this difference is related to the improvements in the low energy
calibration of the Chandra ACIS S3 detector in the interim.

\underline{NE Outer Lobe-Jet}

\begin{figure}[h]
\includegraphics[angle=-90,scale=.22,bbllx=10 pt,bblly=10 pt,bburx=600 pt,bbury=650 pt]{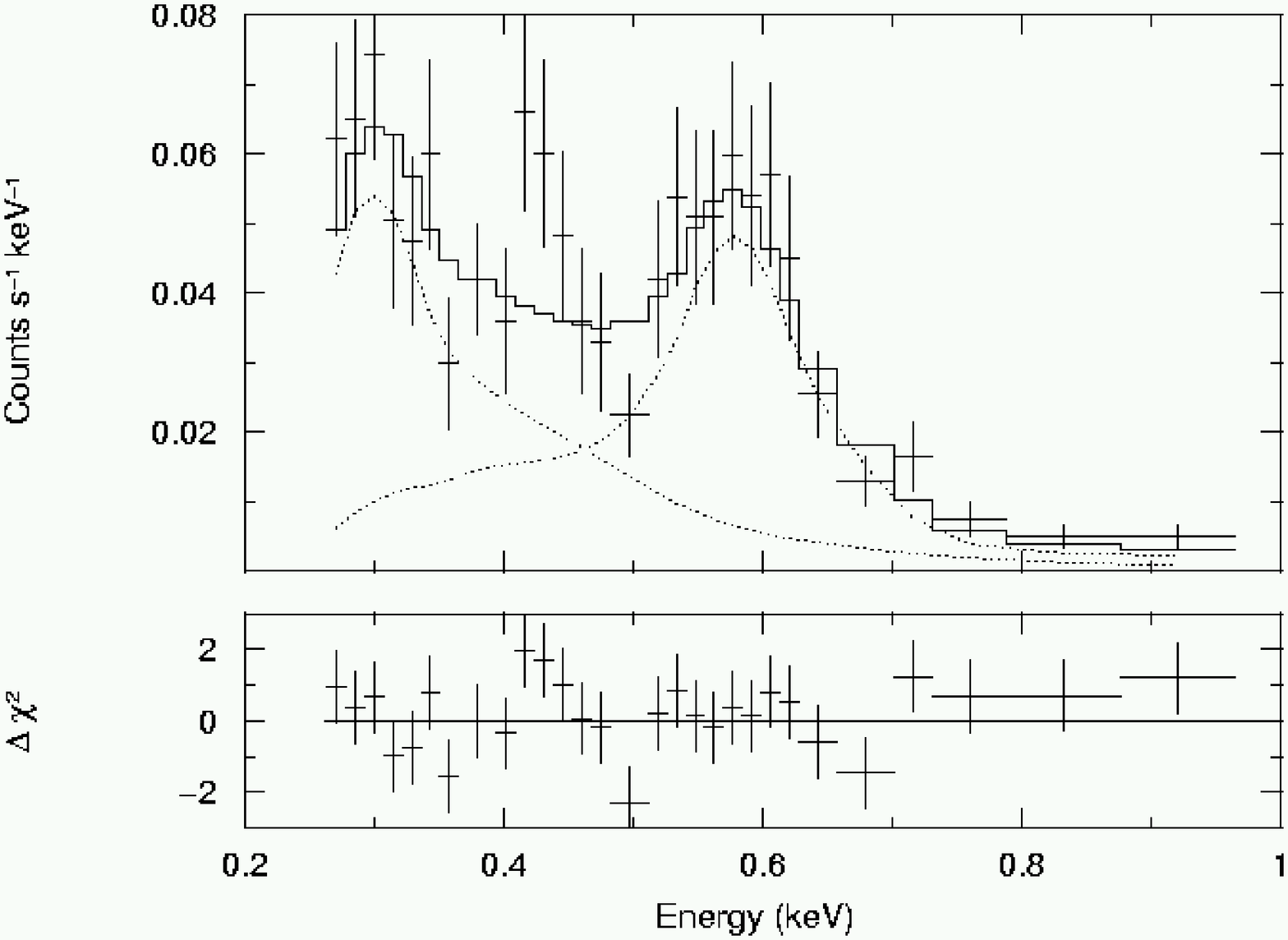} \\
\includegraphics[angle=-90,scale=.22,bbllx=10 pt,bblly=-80 pt,bburx=600 pt,bbury=150 pt]{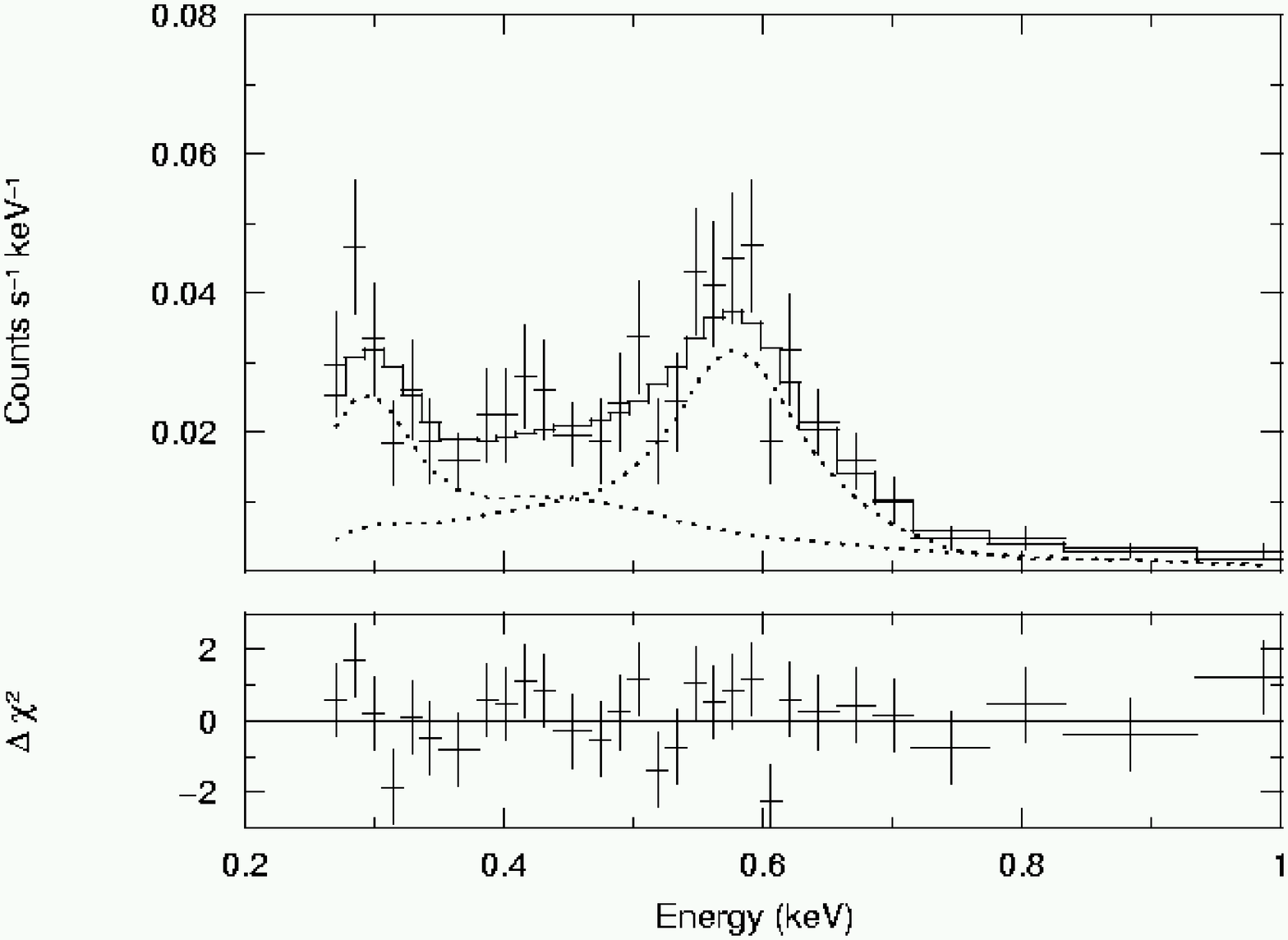}
\caption{X-ray spectrum of NE thermal lobe-jet as observed by {\it Chandra}. The plots show pulse
height spectra and model fits --- 2000.7 above,
2004.0 below. The dotted lines show the contributions from the individual additive components of the model.
Nearby regions of the image with no enhanced emission were used to estimate and subtract the backgound.
 \label{NEoutersub}}
\end{figure}

No spatial variations were detectable in the 2004.0 NE outer lobe-jet
spectrum.  We compared spectra drawn from the half of the outer lobe-jet
closest to the central star and the half farthest away.  We also
compared spectra from an elliptical region centered on the NE outer
lobe-jet bright spot and a surrounding concentric elliptical annulus.  We
found no significant differences in spectral parameters in the pairs
of regions. Therefore, we fitted spectrum models to the entire NE lobe-jet
for the two epochs.  Figure \ref{NEoutersub} shows the spectra of the
NE thermal lobe-jet. In addition to the APEC model with standard abundances,
we had to add a continuum to obtain an acceptable fit. The simplest
continuum is a power law. Adding this component gave an acceptable fit
to the data for both epochs. (Table \ref{netbl}).

There is a hint of a peak in the 2000.7 data at about 0.4 keV, which
we earlier interpreted as N VI emission. However, using the latest
calibrations we can obtain an acceptable fit to the data without
modeling that line.

\underline{SW Outer Lobe-Jet}

\begin{figure}[h]
\includegraphics[angle=-90,scale=.25]{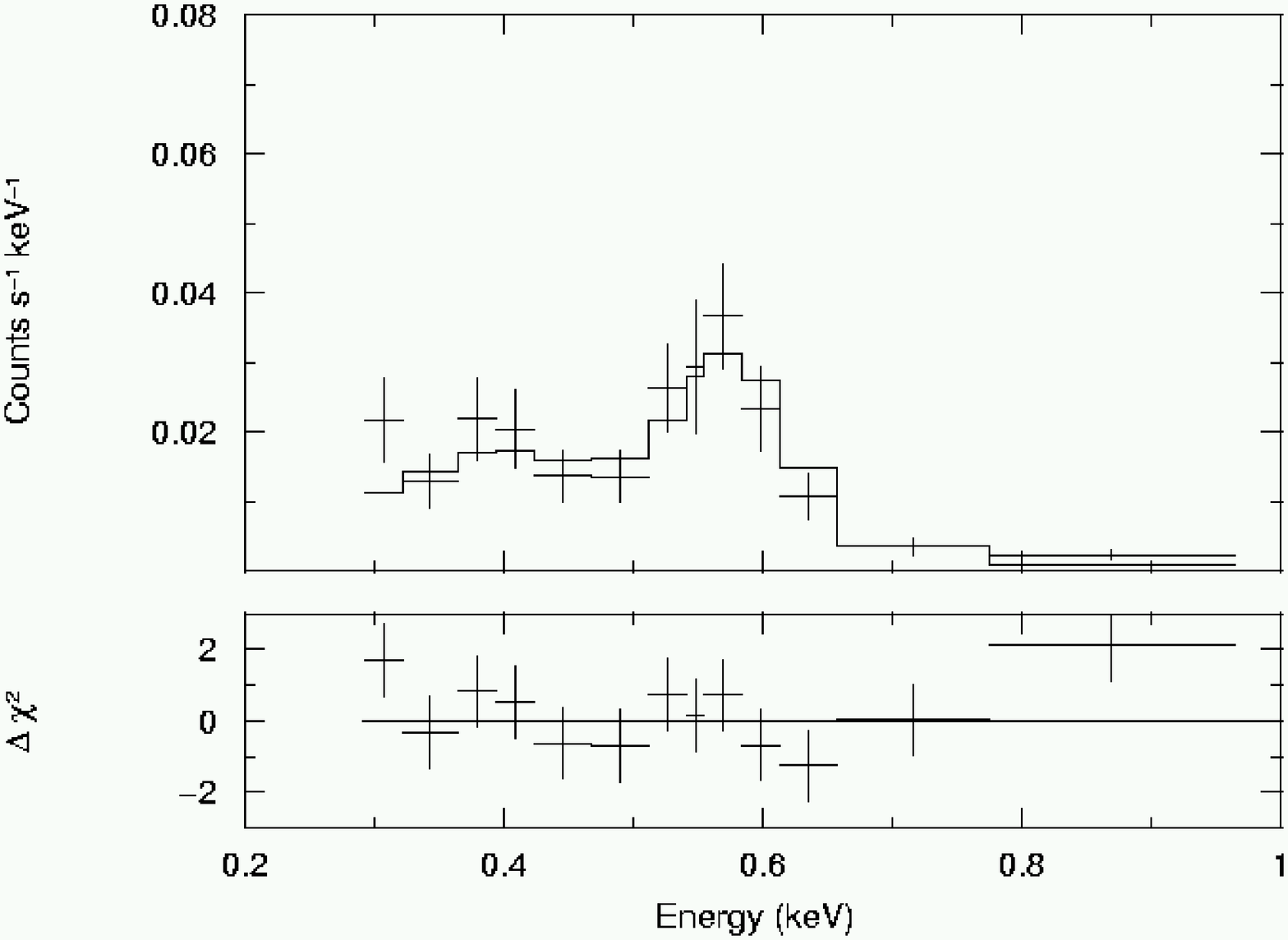} \\
\includegraphics[angle=-90,scale=.25]{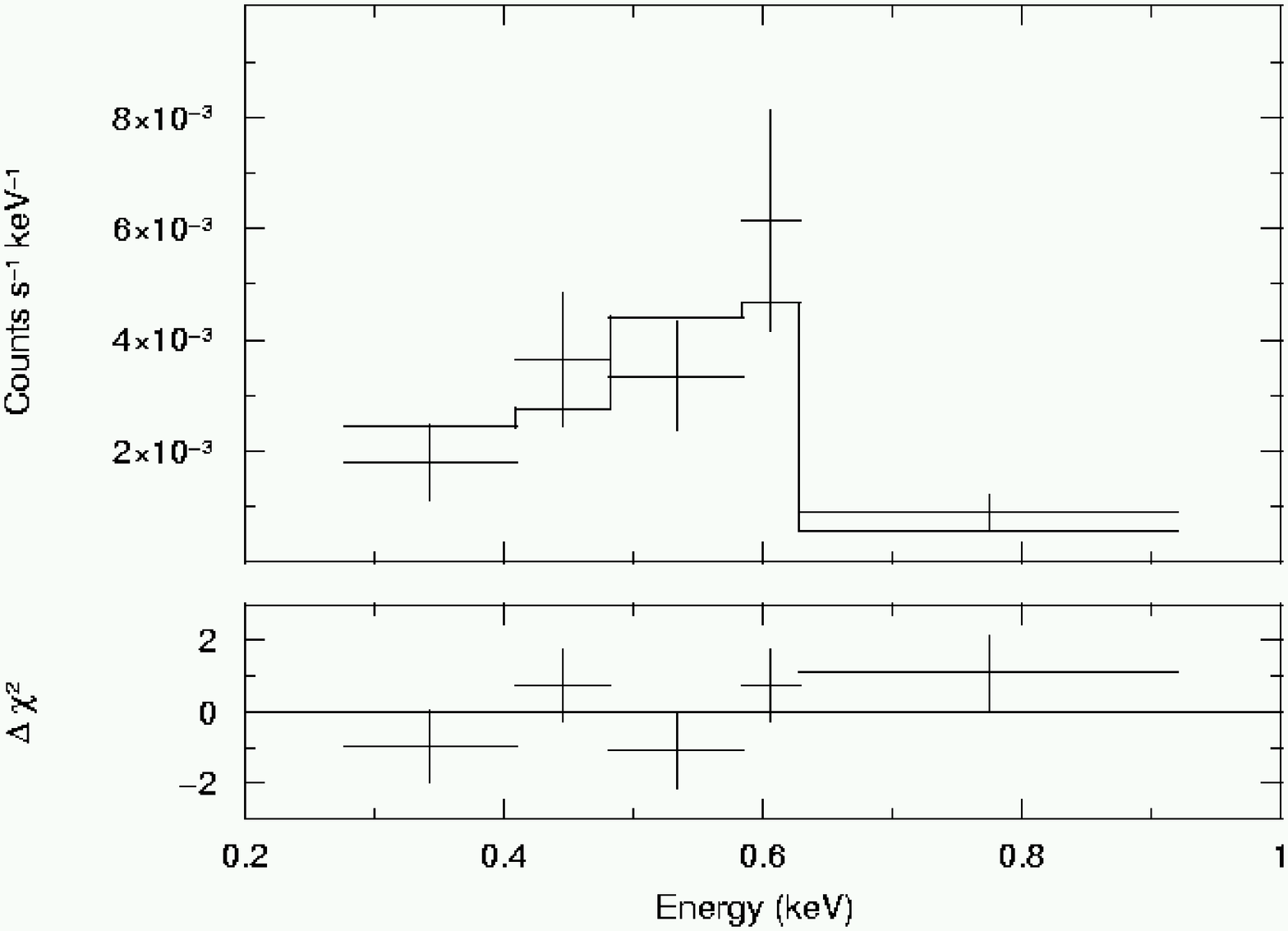}
%%\plottwo{f6l.eps}{f6r.eps}
\caption{X-ray spectra of the SW lobe-jet as observed by {\it Chandra}, 2000.7,
above and  2004.0, below. .\label{SWlobe}}
\end{figure}

Figure \ref{SWlobe} shows the SW outer lobe-jet {\it Chandra} pulse height
spectra in the two epochs, with the model fits. There is a prominent
peak at the energy of the O~VII line in epoch 2000.7, suggesting again
the APEC models as appropriate. In epoch 2004.0, the SW lobe-jet has faded
to about 1/4 the overall intensity.  Unlike for the NE lobe-jet, a power
law was not required for the SW lobe-jet.  The APEC model gives an
acceptable fit. We tested other model fits, including blackbody and
power law instead of APEC, but they gave unacceptable fits.See Table
\ref{swremtbl}. 

\subsection{VLA Observations\label{sec:vlaobs}}

To compare the radio structure of the jets in 2004.0 to the radio
structure in 1999.8 (KPL), we performed VLA observations on 2004 January
6 with the NRAO Very Large Array in the B configuration. The receivers
were tuned to the U (2cm), X (3.5cm), and L (20cm) frequencies with
100 MHz bandpasses. Two scans were made at X band ($\sim$18 minutes
each) and one scan each at U and L bands ($\sim$11 minutes each).  We
used AlPS software for standard calibration, self-calibration, and
imaging.

\begin{figure}[h]
\plotone{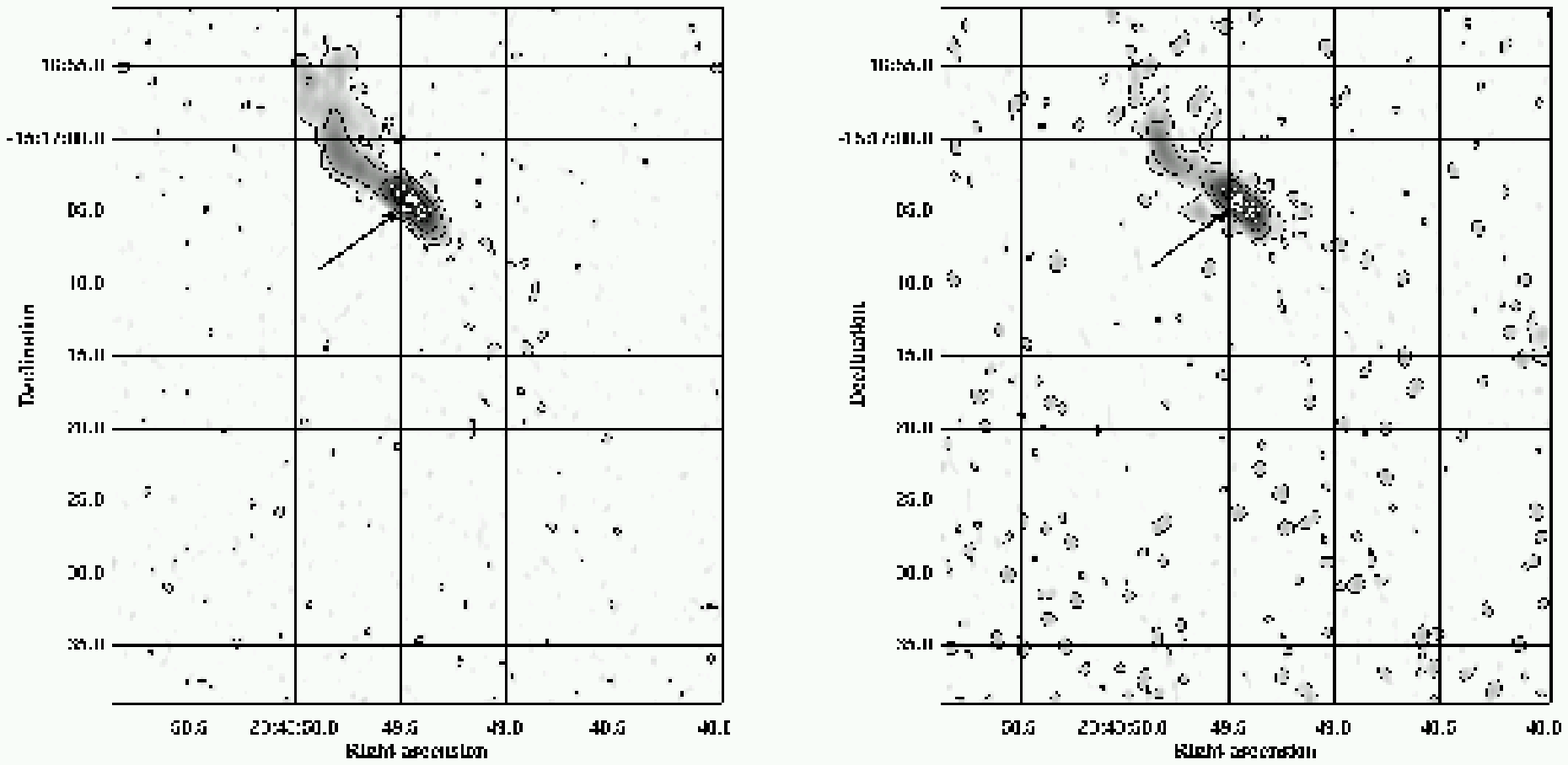}
\caption{VLA radio contour plots of R~Aqr, 1999.8 on left, 2004.0 on
right. The arrows point to the white dots at the position of the central star,
with arrows pointing to them. The gray scale image is the VLA X band
radio intensity, emphasizing only the brightest regions. Contour
levels are 0.015, 0.078, 0.40, 2.10, 4.0 mJy/beam.
\label{VLAChandraCON}}
\end{figure}

Figure \ref{VLAChandraCON} shows the VLA 3.5 cm (X band) observations
of R~Aqr in the two epochs. The beam is $1.0\arcsec\alpha\times 0.69
\arcsec \delta$ PA$\simeq 0\degr$ for both epochs. For the 1999.8
observation, the total flux was 29$\pm0.9$ mJy. For 2004.0 it was
21$\pm0.6$ mJy.  The faint extended radio emission in the outer NE jet
region that was present in 1999.8 has largely disappeared in
2004.0. The brighter wisp in the southern portion of the NE jet is
still present. Its peak has moved east by 0.3$\pm$0.2$\arcsec$ between
1999.8 and 2004.0.  In 2004.0, the peak of the NE jet 3.5 cm emission
is about 6.3$\arcsec$ away from the position of the central binary.
This apparent velocity is several times smaller than the apparent bulk
velocity of the X-ray emitting plasma. There is no detectable emission
at the position of the SW X-ray lobe-jet in either epoch. This motion gives
an apparent bulk velocity of $\sim 90d\pm60d$~km s$^{-1}$, close to or
perhaps consistent with no motion at all.

%\subsection{Composite {\it Chandra} X-ray and VLA radio images}

\begin{figure}[h]
\plotone{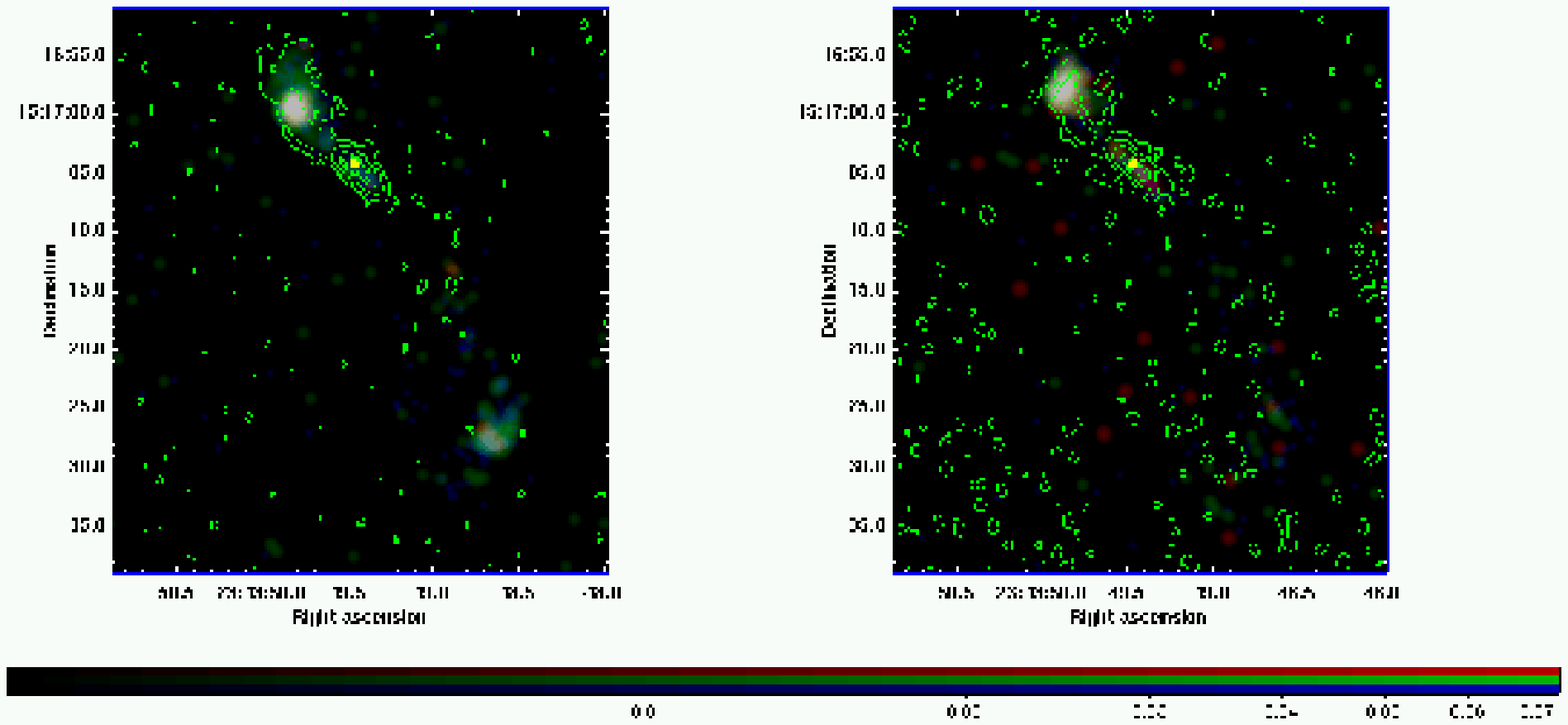}
\caption{Composite VLA radio and {\it Chandra} X-ray images of R~Aqr,
1999.8/2000.7 on left, 2004.0 on right. The VLA 3.5 cm contours are
overlaid on a three color {\it Chandra} X-ray image. Red is the Si~XI/C~VI
energy range, green the N~VI range and blue the O~VII range. The location of R Aqr is indicated by the arrow pointing to the yellow dot. Coordinates are J2000. Color bars are log intensity, in units of counts per 1/8$\arcsec$ pixel.
\label{VLAChandra}}
\end{figure}

Figure \ref{VLAChandra} shows composite images of the {\it Chandra} and VLA
observations from the two epochs, with the VLA 3.5 cm contours
overlaid on a three color {\it Chandra} X-ray image.  The NE X-ray bright
spot lies farther away from the central binary than the NE radio
bright spot at both epochs.  The difference in distances from the
central binary is even larger in 2004.0 than in 2000.7.

\section{Interpretation\label{sec:interp}}

\subsection{Northeast Outer Lobe-Jet}

We find in \S\ref{sec:xrayim} that the peak X-ray emission from the NE
outer lobe-jet has moved $2.0\arcsec\pm 0.1\arcsec$ in 3.3 years. This
motion corresponds to a velocity of at least $v_{NE} = 580d \pm 30d$
km~s$^{-1}$ (at an inclination angle of $20\degr$ as believed for
these lobe-jets \citep{ho91} it becomes $620d$).  

What are the possibilities for the origin of the X-ray emission in the
NE outer lobe-jet? Evidence that could be used to suggest an origin:

\begin{itemize}
\item The thermal or quasi-thermal X-ray spectrum of the NE lobe-jet
\item The radio spectrum of the NE lobe-jet
\item The appearance of new nonthermal jets in the inner region close to the star
\item Motion of the X-ray bright spot over 3.7 years
\item Changes and constancies in the X-ray and radio bright spots over 3.7 years
\item UV and optical observations from past years
\end{itemize}

Completely nonthermal emission mechanisms in this NE lobe-jet are unlikely
because of the shape of the X-ray spectrum. If the spatial position of
the peak of X-ray emission in the NE lobe-jet had not moved between our two
observations, we might think the origin of the thermal component is just a
cloud of hot gas. However, the motion of the X-ray bright spot implies
a high enough velocity to cause a shock, if conditions in the medium
it encounters are appropriate. If we can show that such a shock can
produce the observed emission, then other explanations are
unnecessary. Therefore, we will explore that possibility.

In order to compare our observations with a shock model, we consider
the conditions for the model's validity. First, are we dealing with a
fluid? Is the mean free path for particles in the environment $\ll$
the size of the emitting region? This can be determined from the
density of the environment. We don't know the density directly, and it
is likely to be nonuniform as a result of the complex history of wind
emission from the Mira and its interaction with the compact
companion. We can use evidence from previous optical and UV
observations of the region to get an estimate of the possible range of
densities.  Using \citet{kaf86, me01, ho91}, we obtain n$_e \sim
10^{(2-4)} {\rm cm}^{-3}$. From standard gas kinetic theory, the
Maxwell's mean free path is then $ 3\times 10^{(10-12)}$ cm.  Taking
the X-ray emitting region to be approximately 3$\arcsec \diameter$,
$600d$ AU,  it is $3,000d - 300,000d$ mean free
paths in size. Therefore, the fluid approximation is reasonably
valid. However, it is easy to imagine that electromagnetic phenomena
may also play a role even if mechanical collissions are insufficient.

Most of the work on interpreting astrophysical observations as shocks
has been done for supernova remnants, and most of that has the
objective of predicting the optical and UV line emission for
comparison with the preponderance of optical and UV
observations. Here, we are dealing not with a ``spherical'' blast
wave, but with a directed beam of material, i.e. a jet, and we are
interested in comparing the predictions with X-ray observations. Work
on interpreting jets has concentrated mostly on relativistic jets in
QSOs and NS or BH binaries, or nonrelativistic jets in protostars. We
are dealing with a binary system with a compact star that is
apparently not a NS nor BH, more likely a WD.

However, outflows in HH objects are analoguous to those seen in R
Aqr. The HH jets can be interpreted as a bow shock around an outflow
moving into a medium that itself is moving outward into the ISM at a
different speed \citep{pra01}.  Similar movements in R Aqr can be
derived from the X-ray spectrum and the apparent velocity. The fitted
plasma temperature is $1.74\pm0.23\times$10$^6$ K, corresponding to a shock
velocity of $355\pm47$ km s$^{-1}$ \citep[eqn.~2.24]{dr93}.

At these temperatures, we can consider the plasma as essentially
nonradiative, because the emissivity is very low at such temperatures,
compared with the emissivity at $10^4$ K. This means that we can treat
the problem as adiabatic, and use the Rankine-Hugoniot
equations. Assuming that the preshock medium is no hotter than $10^4$
K, its sound speed is no greater than 17 km s$^{-1}$. At a shock speed
of 355 km s$^{-1}$ its Mach number is $M=21$. This is the regime of
strong shocks, so the relation between the upstream and downstream
densities $\rho_0$ and $\rho_1$ and corresponding velocities $u_0$ and
$u_1$ is $\rho_0 u_0 = \rho_1 u_1$. Since $\rho_1 = 4 \rho_0$ for high
Mach number, $u_1 = u_0/4$. These relations apply in the frame where
the shock is at rest, with the preshock material moving inward toward
the front of the shock and the postshock material emitting the X-rays
and moving outward away from the back of the shock. In this frame,
adopting the convention that positive velocity means motion outward
from the central star, $u_0 = -355 \, {\rm km\, s}^{-1}$ and $u_1 =
-89\, {\rm km\, s}^{-1}$. If we now transform to a frame in which the
preshock medium is at rest by adding 355 km s$^{-1}$ to $u_0$ and
$u_1$, we have $u_0 = 0$, shock velocity $u_{s} = +355
\, {\rm km\, s}^{-1}$ and post-shock velocity $u_{ps} = +266 \, {\rm km\,
s}^{-1}$, the velocity we would expect to observe for the X-ray
emitting region, 3/4 of the unobserved $u_{s}$. However, the velocity
we actually observe for the X-ray emitting region, $580d$ km~s$^{-1}$
(derived in \S\ref{sec:xrayim}) implies a shock velocity in the
Rankine-Hugoniot frame of 4/3 that, or $773d$ km~s$^{-1}$. This
corresponds to a higher post-shock plasma temperature of $8.2\pm0.4
\times 10^6d^{2}$~K, which is incompatible with our observed
spectrum.

The X-ray emission from the outer lobe-jets is thus not consistent with the
temperature we derive from its apparent motion.  Using the shock
scheme of \citet{pra01}, this can be resolved if the shock
is encountering a medium that is already moving in the observer's rest
frame outward ahead of the jet at $314\pm46d$ km s$^{-1}$ so that the
relative motion of the post shock region and the ambient medium is
(3/4 $u_s)=266\pm36$ km s$^{-1}$, with $u_s$ as obtained from the
X-ray spectrum. Incidentally, this implies the unobserved shock is
moving in the observer's rest frame at 670 km s$^{-1}$.

Of course, the temperature of $8.2\times 10^{6}d^{2}$~K derived from
the motion of the X-ray emitting region assumes thermal
equilibrium, which is not always the case at a shock front. For
example, in bow shock models there are ways material can cool by
escaping perpendicular to the direction of motion
\citep{har01}. This leaves the alternate possibility that the ambient material is not moving outward.

In order to characterize the density of the lobe-jets' X-ray emitting
material we use the unabsorbed 0.25 - 2.0 keV flux from the NE outer
lobe-jet (see Table \ref{netbl}) with a resulting luminosity of L$_x$ = 7.0
$\times10^{29}d^{2}$ ergs s$^{-1}$. Taking the X-ray emitting region
to be approximately 3$\arcsec \diameter$, or $600d$ AU, and assuming
the X-ray radiation is optically thin bremsstrahlung emission from a
plasma with solar abundances, the post-shock number density in the NE
outer X-ray lobe-jet is

\begin{equation}\label{lumdensity}
n_{ps}  =  \left( \frac{L_x}{V\Lambda} \right)^{1/2} \\
  =  135\, {\rm cm}^{-3} \left( \frac{f_x}{1.46\times 10^{-13} erg \, cm^{-2}s^{-1}} \right)^{1/2} \\ 
 \left(\frac{\Lambda}{1 \times 10^{-22}\, {\rm erg\, cm}^3\, {\rm s}^{-1}} \right)^{-1/2} d^{-1/2}
\end{equation}

\noindent where $n_{ps}$ is the 2000.7 number density, $f_x$ is the 2000.7 X-ray source flux and $\Lambda$ is the value of the radiative cooling function from \citet{bin87}  at T $\sim~10^{6}$ K. 

The X-ray radiative cooling time can be found using the density and 
the plasma temperature of 1.74 $\times 10^6$ K.  This cooling time,
which takes into account bound-bound, free-free, bound-free,
free-bound and electron scattering radiation \citep{bin87} is 

\begin{equation}\label{tradcool}
t_{rad}=\frac{3}{2}\frac{kT}{n\Lambda} \approx 843\; {\rm yrs} \left(
\frac{T}{1.74\times 10^6\, {\rm K}} \right) \left( \frac{n_{ps}}{135\, {\rm cm}^{-3}} \right)^{-1} \left(
\frac{\Lambda}{1 \times 10^{-22}\, {\rm erg\, cm}^3\, {\rm s}^{-1}}
\right)^{-1}.    
\end{equation}

\noindent This estimated radiative cooling time of the NE lobe-jet is $\sim 843 d^{1/2}$ yrs. The values of source intensity quoted for the two epochs in Table 1 are not different at 90 per cent confidence, so we cannot show any compelling evidence that the NE lobe-jet has changed its X-ray luminosity between the two observations.

The power law component of the NE lobe-jet as seen in the fit to the energy spectrum could be related in some way to the power law emission seen in the inner jets \citep{ni01}.  Perhaps we are even seeing nonthermal emission from the very jet that is heating the lobe gas to million degree temperatures.

\subsection{Southwest Outer Lobe-Jet}

In the SW, the X-ray emission largely disappeared between 2000.7 and
2004.0, presumably because the hot gas cooled to below $\sim
10^{6}$~K.  The roughly 3$\arcsec$ X-ray bright spot corresponds to an
emission region with $600d$ AU diameter, as in the NE lobe-jet.  The X-ray
luminosity in 2000.7, using the unabsorbed 0.25 - 2.0 keV flux listed
in Table
\ref{swremtbl}, was L$_x=1.9\times 10^{29} d^2$ erg s$^{-1}$.  From
equation~(\ref{lumdensity}), the SW outer lobe-jet density is 71$d^{-1/2}$
cm$^{-3}$.  Using this density and Equation~(\ref{tradcool}), the
radiative cooling time is  1288 $d^{1/2}$ years.  Of course, it is
always possible that the emitting region is clumped so that
its average density is much higher and its cooling time much
shorter. However, we see no evidence for that in the {\it Chandra}
image at a size scale of $\ge$100 AU. It could also be that X-ray
radiation is not the only source of cooling for the SW lobe-jet.  The
timescale for cooling by adiabatic expansion could  produce
an observable effect in just a few years.  Assuming the X-ray
emitting region is expanding adiabatically at the sound speed,
%\sim 240 (T/(6 \times 10^6\; {\rm K}))^{1/2}$,
%of 270 km s$^{-1}$ (T/($7~\times 10^6\, {\rm K}))^{1/2}$,
%600 km s$^{-1}$, 
in 3.3 years the radius of the emitting region increases to $R_2 = R_1 + c_s \times t$.
The temperature change can be calculated via the adiabatic relation:
\begin{equation}\label{adiabatic}
(\frac{V_{1}}{V_{2}})^{\gamma-1}=\frac{T_{2}}{T_{1}}.
\end{equation}
Using $\gamma$=5/3 for a monatomic gas, with c$_{s}$=$\sqrt{\frac{\gamma{kT}}{\mu m_{p}}}$=123 km s$^{-1}$ \citep[pg.~163]{os01} the final temperature after adiabatic expansion for 3.3 y is expected to be lower by 40\%. If we thus assume the lobe-jet is expanding adiabatically the plasma temperature is expected to drop from 1.39$\times 10^6$K to $8.4 \times 10^5d^{-1/2}$ K (for small changes in {\em d}) between 2000.7 and 2004.0.  This temperature drop would shift the X-ray emission to lower energies, difficult for Chandra to detect.

\subsection{Mass, Kinetic Energy and Mass Loss in the Lobe-Jets}
Using the density derived above, the images for size scale and the
assumption of a uniform spherical emitter, the mass of the lobe-jets
emitting X-rays in 2000.7 is estimated for the sum of the NE and SW
lobe-jets to be 8.3 $\times$ 10$^{-8}\,d^{5/2}$ M${_\sun}$.  The mass loss
from the giant was estimated to be $3.7\times$10$^{-8}$ M${_\sun}$
yr$^{-1}$ \citep{sea90} (but see \citet{kaf82}, a much higher estimate based on assumption of a high orbital eccentricity), so the X-ray lobe-jet mass is roughly equal to the
total mass loss from the giant in a few years. In the NE lobe-jet we can estimate the mass loss by combining the estimated mass of the x-ray emitting material, its measured outflow velocity and its size to be $\dot{M}_{NE}= 1.1\pm0.2\times 10^{-8}d^{5/2} {\rm M_{\sun} yr^{-1}}$. We cannot make a similar estimate for the SW lobe-jet since it faded so much.

The kinetic energy in the lobe-jet is of interest in order to determine if
it can provide the energy in the lobe-jet to power the X-ray emission.
Using the velocity of 580 {\em d} km s$^{-1}$ derived above, the
kinetic energy in the NE lobe-jet is 1.8$\times$ 10$^{34} d^{9/2}$ J.  The
thermal energy needed for the X-rays at the temperature of
T=1.74$\times$10$^{6}$ K, is 1.9$\times$10$^{33} d^{5/2}$ J. The lobe-jet kinetic energy is about ten times its thermal energy and could power its x-ray emission for 2000-6000 yr.

\section{Conclusions\label{sec:concl}} 

The NE lobe-jet X-ray emission is interpreted as due primarily to a hot plasma witha faint power law component, indicating an energetic collmiated outflow interactin with circumstellar gas at a distance of 1600-2000{\it d} AU from the central binary. Its characteristics appear similar to those of HH objects.

Our calculated value of the NE lobe-jet's density can be compared to
previous determinations \citep{kaf86, me01, ho91}. These previous
reports of density are derived from UV and optical emission line
strength measurements and modeling. They range from n$_e \sim 10^{5}$ in
the HII region surrounding the central star to n$_e \sim 10^{4}$ in
the brighter optical/UV knots, to n$_e \sim 10^{2}$ in a region near
our SW lobe-jet. It appears the density of the X-ray emitting region
is quite different from that in most of the regions emitting UV and
optical. However, we have examined the locations of the regions
emitting UV and optical in these reports and found none to be
coincident in position, and of course none in time with the X-ray
density determinations reported here. The whole of the observations
and modeling to date suggest a region surrounding R Aqr with density
and temperature varying considerably in time and space.  The higher
density of UV emitting material may have accumulated over a longer
time, perhaps being hot enough originally to emit X-rays, then cooling
to UV temperature.  

The mass of the X-ray emitting material is found to be of order the
mass loss per year of the red giant. If the material emitting X-rays
in these lobe-jets is coming from the giant wind, it's unlikely that
more than of order one percent of the wind ends up in the lobe-jets,
because of the large separation between the giant and the compact
companion, and the inefficiency of capture and ejection by the compact
companion. Therefore, this material would have accumulated over
perhaps a few hundred years, unless it was ejected in an episode of
much greater mass loss, such as occurred in 1925 and perhaps again in
the 1964-1974 time frame \citep{wil81} Other symbiotics such as CH Cyg
and Z And have had outbursts of $\sim 10^{-6}{\rm M}{_\sun}$
\citep{hac86, sko02}. Another scenario to account for the mass around the NE lobe-jet is that the X-ray emitting material is not all jet material but surrounding nebula
heated by the jet.  The compact object could eject a small amount of
energetic material in a tightly collimated jet. The jet then impacts
the dense wall \citep{sol01}, ablating material at temperatures high
enough to emit X-rays.

In the NE lobe-jet region the product of density and temperature,
i.e. the pressure, is about the same for the UV emitting regions
($10^4$ cm$^{-3}\times10^4$ K) and the X-ray lobe-jets ($10^2$
cm$^{-3}\times10^6$ K) . However, in the SW the pressure of the UV
emitting material ($10^2$ cm$^{-3}\times10^4$ K) is much lower than of
the X-ray material ($10^2$ cm$^{-3}\times10^6$ K). This permits
expansion of the X-ray emitting material in the SW, whereas in the NE
it would be confined by the UV emitting material.

Radio observations of the outer lobe-jets reveal plasma moving at
$\sim$90 km s$^{-1}$, or even slower, considering the error estimates
on that velocity.  This adds to the complex structure of the jet as
the radio emission is lagging behind the X-ray plasma. It suggests
that the radio source is not closely associated with these lobe-jets,
but may be a relic of past activity emitting thermal radiation at a
lower temperature.

The conditions in R Aqr can be compared with those in another nearby
symbiotic binary, Mira itself (o Ceti). They both have roughly the
same orbital separation, 70 AU for Mira (see \citet{wo04}) and
$\sim$20 AU for R Aqr assuming an orbital period of 44 yr
\citep{wil81} and a system mass of 2 M${_\sun}$, but Mira has no
extensive jets. Its accretion rate onto Mira B is estimated to be 8-30
$\times 10^{-10}{\rm M}{_\sun}\,{\rm yr}^{-1}$ with a total mass loss
from Mira A of 0.4-4 $\times 10^{-7}{\rm M}{_\sun}\,{\rm yr}^{-1}$
\citep{wo04}. Perhaps R Aqr has jets because more of the giant's lost mass finds its way to the compact companion.

The properties of the outer lobe-jets in R Aqr can be compared with
parameters of other symbiotic jets. Fifteen out of about 200
symbiotics are known to have optical or radio jets roughly similar to
the R Aqr inner jets. The only other system known to have an x-ray jet
is CH Cyg. The CHCyg Jet is more similar to the inner jets seen in R
Aqr than the outer lobe/jsets discussed in this paper. There is no
obvious thermal 10$^6$ K extended emission. Since CH Cyg is essentially
unresolved, and at the same distance from us as R Aqr, we conclude
that there's no counterpart to the jets discovered here.  Therefore,
at the present state of observational knowledge, R Aqr is not a
typical symbiotic star jet. In fact it is the only symbiotic observed
to have extended lobe-jets. Future high angular resolution x-ray
observations may discover others, however, since there has been no
systematic survey of that type. 

Mass outflow rates in symbiotics have been estimated using radio
observations and models. A comprehensive study was done \citet{sea90}
resulting in mass loss estimates for 26 symbiotics. They range from
$4\times 10^{-9}$ to $3.7 \times 10^{-5}{\rm M}{_\sun}\,{\rm
yr}^{-1}$. Their estimate for R Aqr is $3.7 \times 10^{-8}{\rm
M}{_\sun}\,{\rm yr}^{-1}$. Therefore, one could say that R Aqr's mass
loss appears to be in the mid range of mass losses in symbiotics, so
it does not appear that the extended lobe-jets of R Aqr are caused by
greater mass loss than in other symbiotics.

The {\it Chandra} X-ray data presented here complement previous UV,
optical, and radio investigations in that they are consistent with an
episodic collimated flow originating in an accretion disk system. The
{\it Chandra} data map similar jet structures to those seen at longer
wavelengths. However, they have revealed new characteristics of the
physics of the jet. The lobe-jets are moving much faster than previously
thought by at least a factor of five. However, the temperature we
observe in the NE lobe-jet is lower than expected for a shock generated by
jet material colliding with stationary material, and suggests that the
jet is moving into material that is itself already moving outward. It
appears the cooling time for the SW lobe-jet is much shorter than expected
from radiative cooling, so it may have cooled by adiabatic expansion
into a less dense surrounding medium in the 3.3 years between the two
observations.

%% Included in this acknowledgments section are examples of the
%% AASTeX hypertext markup commands. Use \url without the optional [HREF]
%% argument when you want to print the url directly in the text. Otherwise,
%% use either \url or \anchor, with the HREF as the first argument and the
%% text to be printed in the second.

\acknowledgments{We are grateful to the staff of the {\it Chandra} X-ray Center for providing the reduced data in a timely fashion, and especially to Eric Mandel and Bill Joye for the ds9 image tool with recent improvements. This work was supported by NASA grant GO4-3050A and contract NAS8-39073. The National Radio Astronomy Observatory is operated by Associated Universities, Inc., under cooperative agreement with the National Science Foundation.  J.~L.~S. was supported by the National Science Foundation.}

Facilities: \facility{CXO(ACIS)}, \facility{VLA}.




\clearpage

\begin{deluxetable}{lll}
\tabletypesize{\scriptsize}
%%\rotate
\tablecaption{NE outer thermal lobe-jet X-ray spectral fit parameters\label{netbl}}
\tablewidth{0pt}
\tablehead{
%%\colhead{} & \colhead{c}{Perimeter} & \colhead{c}{Core}    \\
%%\tableline
\\[-0.07 in]
\colhead{Parameter}&\colhead{2000.7 }&\colhead{2004.0 } \\
 \\[-0.2 in]    }
\startdata

%                            |  2000.7      |2004             
Counts, E$\ge$0.2 keV        &  509         & 500              \\
Exposure, s                  &  22717       & 36523          \\
\tableline
\\[-0.07 in]
APEC kT (keV)      &  0.15[0.13,0.17]  & 0.14 [0.12,0.16]  \\
APEC normalization (10$^{-6}$ph~cm$^{-2}$~s$^{-1}$~keV$^{-1}$) 
                             & 32[22,46] &   38[27,47] \\
Power law photon index, $\Gamma$
                             & 5.6[5.3,5.8]  & 5.1[4.7,5.6] \\
Power law normalization ($\times$10$^{-6}$ph~cm$^{-2}$~s$^{-1}$~keV$^{-1}$) 
                             & 1.6[0.25,1.8]           & 1.8[1.0,2.7]   \\
\tableline
\\[-0.07 in]
Observed 0.25-2.0 keV flux (10$^{-14}$ ergs~cm$^{-2}$~s$^{-1}$) 
                             &  9.7[6.7,13.9]  &     8.1[5.8,10.0]    \\
Source  0.25-2.0 keV flux (10$^{-14}$ ergs~cm$^{-2}$~s$^{-1}$) 
                             & 14.6[10.0,21.0]  &    11.7[8.3,14.5] \\
\tableline
\\[-0.07 in]
Reduced $\chi^{2}$           &  0.98      &           0.96       \\
Confidence (\%)              & 49          &         52         \\
\enddata

%% Text for table notes should follow after the \enddata but before
%% the \end{deluxetable}. Make sure there is at least one \tablenotemark
%% in the table for each \tablenotetext.

\tablecomments{ Values are quoted as best\_fit[lower 90\% limit,upper 90\% limit]. }
%%\tablenotetext{a}{steppar reveals shallow contour, lower confidence interval might be [.2,.35]}

%%{Table \ref{netbl} is published in its entirety in the electronic edition of the {\it Astrophysical Journal}. A portion is shown here for guidance regarding its form and content.}

\end{deluxetable}

%% If you use the table environment, please indicate horizontal rules using
%% Do not put multiple tabular environments within a single table.
%% The optional \label should appear inside the \caption command.

%%SW Remnant
\begin{deluxetable}{lll}
\tabletypesize{\scriptsize}
%%\rotate
\tablecaption{SW outer thermal lobe-jet X-ray spectral fit parameters\label{swremtbl}}
\tablewidth{0pt}
\tablehead{
\colhead{}  &\multicolumn{2}{c}{Epoch}  \\
%%\\[-0.07 in]
\colhead{Parameter }      &\colhead{2000.7 }&\colhead{2004.0 } \\
\\[-0.2 in]          }
\startdata

Counts, E$\ge$0.2 keV        & 226            &  65         \\
Exposure, s                  &  22717       & 36523         \\
\tableline
\\[-0.07 in]
APEC kT (keV)      &  0.12[0.10,0.13]  & 0.12 [0.10,0.16]  \\
APEC normalization (10$^{-5}$ph~cm$^{-2}$~s$^{-1}$~keV$^{-1}$) 
                             & 4.6[3.5,7.0] &   1.2 [0.5, 1.8] \\

\tableline
\\[-0.07 in]
Observed 0.25-2.0 keV flux (10$^{-14}$ ergs~cm$^{-2}$~s$^{-1}$) 
                             &3.0[2.3,4.6]    &0.83[0.33,1.76]\\
%%\tableline
Source 0.25-2.0 keV flux (10$^{-14}$ ergs~cm$^{-2}$~s$^{-1}$) 
                             &4.0[3.0,6.1]   &1.1[0.5,1.7]  \\
\tableline
\\[-0.07 in]
Reduced $\chi^{2}$           & 1.07            &1.12      \\
Confidence (\%)              &38              &34        \\
\enddata

%% Text for table notes should follow after the \enddata but before
%% the \end{deluxetable}. Make sure there is at least one \tablenotemark
%% in the table for each \tablenotetext.

%%{Table \ref{swremtbl} is published in its entirety in the electronic edition of the {\it Astrophysical Journal}. A portion is shown here for guidance regarding its form and content.}

\tablenotetext{a}{Parameter fixed}
%%\tablenotetext{b}{Another sample footnote for table~\ref{swremtbl}}
\tablecomments{ Values are quoted as best\_fit[lower 90\% limit,upper 90\% limit]}

\end{deluxetable}

\clearpage

\end{document}